\documentclass[twocolumn,showpacs,preprintnumbers,amsmath,amssymb,prl]{revtex4}

\usepackage{graphicx}
\usepackage{dcolumn}
\usepackage{epsf}

\begin{document}

\title{Intrinsic Josephson effect and nonequilibrium soliton structures in two-gap superconductors}
\author{A. Gurevich$^{1}$ and V.M. Vinokur$^{2}$.}
\affiliation{$^{1}$Applied Superconductivity Center, University of Wisconsin, Madison, Wisconsin 53706}
\affiliation{$^{2}$ Materials Science Division, Argonne National Laboratory, Argonne, Illinois.}

\date{\today}

\begin{abstract}
We predict a new dynamic state in current-carrying superconductors with multicomponent order parameter.
If the current density J exceeds a critical value $J_t$, an interband breakdown
caused by charge imbalance of nonequilibrium quasiparticles occurs.  For $J>J_t$,
the electric field penetrating from current leads gives rise to various static and dynamic
soliton phase textures, and voltage oscillations similar to the nonstationary Josephson effect.  
We propose experiments to observe these effects which would probe the multicomponent 
nature of the superconducting order parameter.

\end{abstract}
]

\maketitle

There are experimental evidences that several superconductors, including
$MgB_2$ \cite{maz,2gap}, $NbSe_2$ \cite{nbse2}, the heavy-fermion $UPt_3$ \cite{upt3}, 
the organic $(TMTSF)_2X$ \cite{gerome} and $\kappa-BEDT$ \cite{org} compounds may have a  
multicomponent order parameter $\psi$ with internal degrees of freedom \cite{legget,su}. 
For instance, for two weakly coupled s-wave order parameters
$\psi_1=\Delta_1e^{i\theta_1}$ and $\psi_2=\Delta_2e^{i\theta_2}$
corresponding to two different parts of the Fermi surface (like in $MgB_2$), the internal
degree of freedom is the interband phase difference $\theta({\bf r},t)=\theta_1-\theta_2$. In this case,  
besides the usual phase-locked states ($\theta=0$ or $\pi$), intrinsic phase textures $\theta({\bf r},t)$ 
can occur. The soft modes associated with the
fluctuations of $\theta({\bf r},t)$ are nearly decoupled from the
gap fluctuations and behave like the Anderson plasmons in
Josephson junctions \cite{legget}. These modes may manifest
themselves as resonances in the ac Josephson effect \cite{agt}, 
or static $2\pi$ phase kinks in $\theta(x)$ \cite{tanaka}.

In this Letter we show that, although static metastable textures $\theta ({\bf r})$ do not manifest 
themselves in the equilibrium magnetic response,   
the situation radically changes for {\it nonequilibrium} current states where     
the charge imbalance near normal leads gives rise to dynamic phase slip structures $\theta({\bf r},t)$ 
propagating into a superconductor (Fig. 1). We predict a new dynamic state above 
the critical current density $J>J_t$ which marks the onset of the current-induced breakdown of
the superconducting state. For the multicomponent $\psi$, such breakdown is a two-stage process.  First, at
$J=J_t$, an interband phase breakdown occurs, resulting in
spontaneous phase solitons in $\theta(x,t)$ and ac voltage oscillations. At higher currents, 
$J=J_d$, both gaps $\Delta_{1,2}$ are suppressed by the pairbreaking effects. 
For weak interband coupling, $J_t$ is much smaller than the depairing current density $J_d$.

We obtain the equations of motion for $\theta$ and the electric field ${\bf E}$ near the critical
temperature $T\approx T_c$, using the time-dependent
Ginzburg-Landau (TDGL) equations \cite{kopnin} generalized to   
a two-gap superconductor:
    \begin{equation}		
    \Gamma_\mu (\partial_t\psi_\mu-2\pi ci\varphi/\phi_0)=
    -\delta F/\delta\psi_\mu,
    \label{fder}
    \end{equation}
where $\mu$ runs from 1 to 2,  $\varphi$ is the electric
potential, $\phi_0$ is the flux quantum, $c$ is the speed of light, $\Gamma_\mu$ are damping
constants,  and the free energy $F=\int d^3{\bf r}(f_1+f_2+f_m+f_{int})$ contains the magnetic part 
$f_m=|\nabla\times\bf A|^2/8\pi$, the GL intraband part $f_\mu$, and the interband interaction $f_{int}$ 
    \begin{eqnarray}
    f_\mu=\alpha_\mu|\psi_\mu|^2+\frac{\beta_\mu}{2}|\psi_\mu|^4
    +g_\mu\left\vert\left(\nabla+\frac{2\pi i}{\phi_0}\bf{A}\right)\psi_\mu\right\vert^2,
    \label{gl} \\			
    f_{int}=\gamma(\psi_1\psi_2^*+\psi_1^*\psi_2)=2\gamma\Delta_1\Delta_2\cos\theta. \qquad
    \label{int}			
    \end{eqnarray}
Here ${\bf A}$ is the vector potential, $\gamma$ is a coupling
constant, the asterisk means complex conjugation. The qualitative results of this work remain valid  
for any other periodic dependencies of $f_{int}(\theta)$ on $\theta$, for example,
$f_{int}=2\Delta_1^2\Delta_2^2(\gamma_1\cos 2\theta+\gamma_2)$,
suggested for heavy-fermion superconductors \cite{upt3}. We focus here on 
two-gap superconductors for which two cases are possible: (i) The Cooper instability occurs at the
same temperature in both bands, thus $\alpha_\mu(T)\propto
T-T_c$, $\gamma\propto T-T_c$, while $g_\mu$, and $\beta_\mu$ are
constants.  (ii) The critical temperature $T_{c1}$
in the band 1 is higher than $T_{c2}$, in the band 2, so $\psi_1$ appears spontaneously
below $T_c$, while $\psi_2$ at $T_{c1}<T<T_c$ is induced by the
interband interaction, because $\alpha_1\propto
T-T_{c1}$, $\alpha_2\propto T-T_{c2}$, and $\gamma$, $g_\mu$ and
$\beta_\mu$ are constants \cite{suhl}. In this paper we consider the first scenario, assuming
that $\gamma\ll\alpha_{1,2}$, which is likely to occur in $MgB_2$ \cite{maz,2gap}. 

Eqs. (\ref{fder}) yield the following gap equations
    \begin{equation}		
    \Gamma_\mu\dot{\Delta}_\mu=g_\mu\nabla^2\Delta_\mu-\tilde{\alpha}_\mu\Delta_\mu-\beta_\mu\Delta_\mu^3-
    \gamma\Delta_{\overline{\mu}}\cos\theta,
    \label{r}
    \end{equation}
where $\overline{\mu}=2$ if $\mu = 1$, and $\overline{\mu}=1$ if $\mu = 2$, $\theta =\theta_1-\theta_2$,
$\tilde{\alpha}_\mu=\alpha_\mu+(2\pi Q_\mu/\phi_0)^2g_\mu$, and
${\bf Q}_\mu={\bf A}-\phi_0\nabla\theta_\mu /2\pi$. The imaginary part of Eq. (\ref{fder}) gives
    \begin{eqnarray}
    \Gamma_\mu\bigl(\dot{\theta}_\mu-\frac{2\pi c}{\phi_0}\varphi\bigr)\Delta_\mu^2=\!
    -\frac{2\pi g_\mu}{\phi_0}\nabla(\Delta_\mu^2{\bf Q}_\mu)\pm\gamma\Delta_1\Delta_2\sin\theta
    \label{t}\\			
    \nabla\times\nabla\times{\bf A}=(4\pi/c)(\sigma{\bf E}+{\bf J}_s)\qquad\qquad
    \label{max}			
    \end{eqnarray}
with the plus sign corresponding to $\mu=1$. The current
density ${\bf J}=\sigma{\bf E}+{\bf J}_s$ has the ohmic
component ${\bf J}_n=\sigma{\bf E}$ proportional to the electric
field ${\bf E}=-\nabla\varphi-\dot{\bf A}/c$, and
the normal conductivity $\sigma$. The supercurrent
    \begin{equation}		
    {\bf J}_s=-8\pi^2c(g_1\Delta_1^2{\bf Q}_1+g_2\Delta_2^2{\bf Q}_2)/\phi_0^2
    \label{js}
    \end{equation}
is a sum of independent intraband contributions. The boundary
conditions, $(i\partial_n+2\pi A_n/\phi_0)\psi_\mu=i\psi_\mu/l_\mu$, 
between a superconductor and a normal metal
imply zero perpendicular supercurrents for {\it both} $\psi_1$ and
$\psi_2$. Excluding ${\bf Q}_{1,2}=-(\phi_0{\bf J}_s \pm
4\pi c g_{2,1}\Delta_{2,1}^2\nabla\theta
)\phi_0/[8\pi^2c(g_1\Delta_1^2+g_2\Delta_2^2)]$ from Eqs. (\ref{gl}) and
(\ref{js}), we obtain the energy density
$f=f_\theta+f_e+\sum_\mu[\alpha_\mu\Delta_\mu^2
+g_\mu(\nabla\Delta_\mu)^2+\beta_\mu\Delta_\mu^4/2]$, where the magnetic 
and phase dependent parts, $f_m$ and $f_\theta$ turn out to be decoupled from each other:
    \begin{eqnarray}
    f_m=\frac{H^2}{8\pi}+\frac{\phi_0^2J_s^2}{(4\pi c)^2(g_1\Delta_1^2+g_2\Delta_2^2)}
    \label{fe}\\			
    f_\theta=\frac{g_1g_2\Delta_1^2\Delta_2^2(\nabla\theta)^2}{(g_1\Delta_1^2+g_2\Delta_2^2)}
    +2\gamma\Delta_1\Delta_2\cos\theta
    \label{ft}			
    \end{eqnarray}
For constant gaps $\Delta_\mu$, variation of Eq. (\ref{fe}) with respect to
${\bf H}$ yields the London equation $\lambda^2\nabla^2{\bf
H}=\bf{H}$ with the magnetic penetration depth
$\lambda^2=\phi_0^2/32\pi^3(g_1\Delta_1^2+g_2\Delta_2^2)$.  If 
$\gamma\ll\alpha_{1,2}$, the gaps $\Delta_\mu$ are
decoupled from $\theta$, so variation of Eq. (\ref{ft}) with
respect to $\theta$ yields the sine-Gordon equation
$L_\theta^2\nabla^2\theta=\pm\sin\theta$ which has a single-soliton
solution \cite{tanaka}, or periodic solutions analogous to the 
Josephson vortices in a magnetic field \cite{barone}.
However, the interband phase difference $\theta$ 
is {\it decoupled} from static magnetic fields, while the Gibbs free
energy of a Josephson contact contains the interaction term
$\propto (\bf H\nabla\theta_J)$. Thus, equilibrium nonuniform
solutions $\theta(x)$ are energetically unfavorable as compared
to the uniform phase-locked state $\theta=0$ for $\gamma<0$ or
$\theta=\pi$ for $\gamma>0$. By contrast, various dynamic or quenched 
phase textures can be generated during current-induced interband breakdown.   

We consider the weak coupling limit $\gamma\ll\alpha_\mu$, for which the uniform 
gaps $\Delta_{1,2}$ are unaffected by $\theta({\bf r},t)$, so Eqs. (\ref{t})-(\ref{max}) give a time-dependent London equation
with the account of dissipation and charge imbalance effects \cite{imb}. 
Expressing ${\bf Q}_{1,2}$ in Eqs. (\ref{t}) in terms of ${\bf J}$ and
$\nabla\theta$, and then subtracting the equations for $\theta_1$ and $\theta_2$ 
from each other, we obtain the following equation for $\theta$
    \begin{equation}		
    \tau_\theta\dot{\theta}=L_\theta^2\nabla^2\theta\pm\sin\theta+\alpha_\theta\mbox{div}{\bf J}_s,
    \label{theta}
    \end{equation}
where the relaxation time $\tau_\theta$, the decay length $L_\theta$, and
the charge coupling parameter $\alpha_\theta$ are given by
    \begin{eqnarray}
    \tau_\theta=\Gamma_1\Gamma_2\Delta_1\Delta_2/|\gamma|(\Gamma_1\Delta_1^2+\Gamma_2\Delta_2^2),
    \label{tt} \\			
    L_\theta^2=g_1g_2\Delta_1\Delta_2/|\gamma|(g_1\Delta_1^2+g_2\Delta_2^2),
    \label{tl} \\			
    \alpha_\theta=\frac{\phi_0\Delta_1\Delta_2(\Gamma_1g_2-\Gamma_2g_1)}
    {4\pi c|\gamma|(g_1\Delta_1^2+g_2\Delta_2^2)(\Gamma_1\Delta_1^2+\Gamma_2\Delta_2^2)}
    \label{alp}			
    \end{eqnarray}
The signs in Eq. (\ref{theta}) correspond to the sign of $\gamma$. As seen from
 Eq. (\ref{theta}), the phase mode is unaffected by any distribution
of bulk supercurrents, unless there is a nonequilibrium charge imbalance
$\mbox{div}{\bf J}_s=-\sigma\mbox{div}{\bf E}$ caused by the
electric field penetrating from normal current leads. The excess charge density provides a driving term in 
the sine-Gordon equation for $\theta$, where the coupling constant $\alpha_\theta$ is nonzero if $g_1\Gamma_2\neq g_2\Gamma_1$, 
which implies different electron diffusivities in the bands 1 and 2.  

To obtain the equation for ${\bf E}$, we add Eqs. (\ref{t}) for $\theta_1$ and $\theta_2$, then take the gradient of the sum and
express ${\bf Q}_{1,2}$ in terms of ${\bf J}$ and $\nabla\theta$. This yields
    \begin{equation}		
    \tau_e\dot{{\bf E}}+{\bf E}-L_e^2\mbox{grad}\mbox{div}{\bf E}+\alpha_e\nabla\dot{\theta}=\tau_e\dot{{\bf J}}/\sigma,
    \label{e} 
    \end{equation}
where ${\bf J}(t)$ is the driving current density, $L_e$ is the electric field penetration depth, 
$\tau_e$ is the charging time constant, and the coupling term $\alpha_e\nabla\dot{\theta}$ 
describes an electric field caused by moving phase textures:
 \begin{eqnarray}
    \tau_e=\sigma\phi_0^2/8\pi^2c^2(g_1\Delta_1^2+g_2\Delta_2^2),
    \label{te}\\			
    L_e^2=\sigma\phi_0^2/8\pi^2c^2(\Gamma_1\Delta_1^2+\Gamma_2\Delta_2^2),
    \label{le} \\			
    \alpha_e=2|\gamma|\Delta_1\Delta_2\alpha_\theta\qquad\qquad
    \label{ae}			
    \end{eqnarray}

We first use Eqs. (\ref{theta}) and (\ref{e}) to calculate the phase textures 
in a microbridge of length $2a$ (Fig. 1a) for which Eq. (\ref{theta}) at $\gamma<0$ takes the form
    \begin{equation}		
    L_\theta^2\theta''-\sin\theta+\alpha_\theta J_s'=0
    \label{td}
    \end{equation}
Here $\theta'(x)$ and $J_s(x)$ vanishes at the bridge edges, $x=\pm a$, where 
$E(x,t)=E_0\cosh(x/L_e)/\cosh(a/L_e)$ equals the value $E_0=J/\sigma$ provided by an external power source, 
$\theta'(0)=E'(0)=0$, and the prime denotes differentiation over x. 
Now we obtain the condition under which Eq. (\ref{td}) has
a stable solution $\theta(x,\theta_a)$ localized at the edge of
a long ($a\gg L_\theta$) bridge, where $\theta_a=\theta(a)$. 
We first consider the limit $L_\theta\gg L_e$, for which $J_s'(x)=-\sigma E'(x)$ is essential only in a narrow 
region $a-L_e< x <a$. Multiplying Eq. (\ref{td}) by $\theta'$
and integrating from 0 to $a-L_e$, we obtain  $\theta'(a-L_e) = -2\sin\theta_a/2$. 
Here $\theta'(a-L_e)=-\alpha_\theta J/L_\theta^2$ is obtained by integrating 
Eq. (\ref{td}) over $a-L_e<x<a$, where $\theta(x-L_e)\approx \theta_a$, and 
$\sin\theta$ can be neglected. Excluding $\theta'(a-L_e)$, we arrive at the equation
$\alpha_\theta J=2L_\theta|\sin\theta_a/2|$, which has solutions for $\theta_a$ only below the critical current density
$J_t=2L_\theta/\alpha_\theta$:
    \begin{equation}		
    J_t=\frac{8\pi c(\Gamma_1\Delta_1^2+\Gamma_2\Delta_2^2)\sqrt{|\gamma|g_1g_2(g_1\Delta_1^2+g_2\Delta_2^2)}}
    {\phi_0\sqrt{\Delta_1\Delta_2}|g_1\Gamma_2-g_2\Gamma_1|}
    \label{jcc}
    \end{equation}

In the opposite limit $L_\theta\ll L_e$, the Laplacian in Eq. (\ref{theta}) can be neglected. Then
the steady-state solutions $\sin\theta(x)=-\sigma\alpha_eE'(x)$,
and $E(x)=E_0\cosh(x/L_e)/\cosh(a/L_e)$, exist only if $\sigma\alpha_eE'(a)<1$,
or $J<J_t=L_e/\alpha_e\tanh(a/L_e)$, where
    \begin{equation}		
    J_t=\frac{|\gamma|[\sigma(\Gamma_1\Delta_1^2+\Gamma_2\Delta_2^2)]^{1/2}(g_1\Delta_1^2+g_2\Delta_2^2)}
    {\sqrt{2}\Delta_1\Delta_2|g_1\Gamma_2-g_2\Gamma_1|\tanh(a/L_e)}
    \label{jc}
    \end{equation}
If $\gamma(T)$ and $\alpha_\mu(T)$ linearly pass through zero at $T_c$, Eqs. (\ref{jcc}) and (\ref{jc}) 
yield $J_t\propto (T_c-T)^{3/2}$.

For $J>J_t$, the charge-induced interband breakdown gives rise to
a striking dynamic phase texture in which solitons periodically
appear near the current leads and then propagate to the bulk. The actual
soliton dynamics depends on the particular boundary
conditions, so we performed numerical simulations of Eqs.
(\ref{theta})-(\ref{ae}) for the different geometries shown in
Fig. 1. We consider the limit
$\alpha_e\alpha_\theta\ll\mbox{min}(\tau_\theta L_e^2,
\tau_eL_\theta^2)$ of weak coupling between $\theta$ and $E$,
for which the driving term div{\bf E} in Eq. (\ref{theta}) is
mostly determined by the {\it static} $E(x)$ near the current
lead, while the last term in Eq. (\ref{ae}) gives only a small ac
correction to $E$ of the second order in $\alpha_e$\cite{stat}.

Fig. 2 shows the soliton formation near the current lead in the thin microbridge (Fig 1a)
as J was instantaneously turned on from zero to a value below $J_t$. Such current step
produces a phase soliton localized near the edge, while the bulk of the bridge remains in the phase-locked state
with either $\theta=0$ or $\pi$, depending on the sign of $\gamma$. For $J<J_t$, this behavior is characteristic of all geometries
shown in Fig. 1, whereas the dynamic phase texture for $J>J_t$ can be very different.

We start with the bridge geometry (Fig. 1a), for which $E(x,t)$
and $\theta(x,t)$ are even and odd functions of x, respectively,
so the boundary conditions are: $E(\pm a,t)=E_0$, $E'(0,t)=0$,
$\theta(0)=0$, and $\theta'(\pm a,t)=0$. The latter condition
ensures that supercurrents in both bands vanish at the normal
electrodes, where $J=\sigma E_0$.  In this case a soliton
first appears at the bridge edge, but for $J>J_t$, it is pushed
to the bulk by the strong gradient of $E(x)$. Then the next
soliton forms near the edge and the process repeats periodically,
resulting in the propagation of the soliton chain into the bulk, as shown in
Fig. 3. After the first soliton reaches the center of the bridge,
it stops (because $\theta(0)=0$), while new solitons keep
entering the bridge from the current lead. During this soliton pileup, the mean
slope $\bar{\theta}'(t)$ increases, reaching a critical value
$\bar{\theta}'$ at which the soliton generation at the edge stops
and a static phase texture  forms.  For $J\gg J_c$, the maximum soliton
density $n=\bar{\theta}_c'/2\pi$ can be estimated from the static
Eq. (\ref{theta}) in which the rapidly oscillating $\sin\theta$
can be neglected, and $E(x)=J\sinh(x/L_e)/\sigma\sinh(a/L_e)$.
This yields $\theta'(x)=\alpha_\theta
JL_e[\cosh(a/L_e)-\cosh(x/L_e)]/L_\theta^2\sinh(a/L_e)$, giving
for the long bridge $(a\gg L_e)$:
    \begin{equation}		
    \bar{\theta}_c'=JL_e\phi_0(\Gamma_1g_2-\Gamma_2g_1)/4\pi cg_1g_2(\Gamma_1\Delta_1^2+\Gamma_2\Delta_2^2)
    \label{n}
    \end{equation}
During the formation of the soliton chain, $t<t_c\sim \tau_\theta a\bar{\theta}'_c/2\pi$,
voltage oscillations are generated on the bridge.
A similar transient behavior occurs at the point contact (Fig. 1b), for which div${\bf E}=2I\exp(-r/L_e)r_0/\pi L_e(r_0+L_e)r$,
where $I$ is the current through the semi-spherical contact of radius $r_0$. In this case the increase of $I$ above
$I_c$ causes propagation of concentric soliton shells until the static structure forms as the
critical phase gradient $\sim \theta'_c$ is reached.

A very different soliton dynamics occurs in the 4-terminal geometry (Fig. 1c), for which the supercurrents 
make $90^{\circ}$ turns around the central stagnation point $(x=0)$ where $\nabla\theta=0$. As a result, 
$E(x)$ and $\theta(x,t)$ become odd and even functions of x, respectively, which   
radically changes the dynamics of the phase textures as compared to the bridge, since now the solitons  
do not stop as they reach the center, but keep moving until they reach the opposite current lead. This is due to the fact that 
the driving charge density div${\bf E}$ does not change sign all the way along the horizontal leg of the cross in 
Fig. 1c. The corresponding total extra charge (which plays the same role here as the transport current in a Josephson junction) 
is exactly compensated by the opposite charge distributed along the vertical leg. Thus, the mean phase gradient $\bar{\theta}'$ 
remains constant, while the phase slippage 
at the normal leads results in a continuous soliton shuttle and periodic voltage oscillations
between the current lead and the center as shown in Fig. 4. For $L_e\gg L_\theta$, the ac voltage
$V_\omega(t)=\sum_{m=1}^{\infty}V_m\cos (m\omega t+\psi_m)$
can be estimated by integrating Eq. (\ref{e}) from $0$ to $a$, which yields
$\tau_e{\dot V}_\omega+V_\omega=\alpha_e[{\dot\theta_a(t)-\dot\theta(0,t)}]$.
Here the values $\theta_a(t)\sim\theta(0,t)$ obey the RSJ equation 
$\tau_\theta{\dot\theta_a}+\sin\theta_a=\beta$, for which 
$\tau_\theta{\dot\theta_a}=\sqrt{\beta^2-1}[1+2\sum_{m=1}^{\infty}(\beta-\sqrt{\beta^2-1})^m\cos m\omega t]$, 
$\beta=J/J_t$, and $\omega=\tau_\theta^{-1}\sqrt{\beta^2-1}$ \cite{al}. As a result, 
we obtain $\tan\psi_m=m\omega\tau_\theta$, and
    \begin{equation}
    V_m\simeq \frac{2\alpha_e}{\tau_\theta}\sqrt{\beta^2-1}\frac{(\beta-\sqrt{\beta^2-1})^m}
    {\sqrt{1+(m\omega\tau_e)^2}}.
    \label{v}
    \end{equation}
$V_m(\beta)$ is maximum at $\beta=\beta_m$, where $\beta_m\simeq 1+0.5(\tau_\theta/m\tau_e)^{2/3}$
for $\tau_\theta\ll \tau_e$. Thus, a two-gap {\it superconducting} bridge between normal banks exhibits
the behavior of an overdamped Josephson junction. If the widths of the 
superconducting legs in Fig. 1c are different, either a unidirectional and mutually orthogonal soliton motion  
may occur, if $J>J_t$ in one or both legs, respectively. 

The soft interband phase mode could also manifest itself in
rf absorption at frequencies below the small gap $\Delta_2$. Interaction of the $\theta$-mode with 
the rf electric field depends on the polarization of
${\bf E}$: if ${\bf E}(t)$ is parallel to the sample surface, then div${\bf E}=0$, so
the phase mode is not excited by the rf field. If, however the rf field has a component
perpendicular to the sample surface, then the phase mode contributes
to the rf impedance.

In conclusion, we predict an interband breakdown induced by
nonequilibrium quasiparticles in two-gap superconductors. It results in spontaneous generation of 
static and/or dynamic phase textures, causing the Josephson
voltage oscillations. The observation of these oscillations would unambiguously indicate the
multicomponent nature of the superconducting order parameter.

This work was supported by the NSF  MRSEC (DMR 9214707), AFOSR
MURI (F49620-01-1-0464) (AG), and by the US DOE Office of Science
under contract No. W31-109-ENG-38 (AG and VV).

\clearpage

\begin{figure}          
\vskip \baselineskip
\caption{Geometries in which the interband phase breakdown could occur. Here N 
labels normal electrodes, gray domains show
phase solitons moving along thin arrows, and block arrows indicate current directions.
Static phase textures form in microbridges (a) and point contacts (b), while in the four-terminal geometry (c)
the solitons continuously move from one current lead to another.}
\label{fig.1}
\end{figure}

\begin{figure}          
\vskip \baselineskip
\caption{Formation of a static phase soliton in $\theta(x)$ near the bridge edge after the current
density  was instantaneously turned on from $0$ to $J=0.99J_t$ at $t=0$. Times and distances 
from the center $(x=0)$ are taken in the units of $\tau_\theta$ and a, respectively, $L_e=a/10$, $L_\theta=0.1L_e$.}
\label{fig.2}
\end{figure}

\begin{figure}          
\vskip \baselineskip
\caption{Dynamics of formation of a static soliton chain in the bridge of length 2a after
$J(t)$ was instantaneously turned on from $0$ to $1.025J_t$ at $t=0$. 
Only the right half $(0<x<a)$ is shown, and the rest is the same as in Fig. 2.}
\label{fig.3}
\end{figure}

\begin{figure}          
\vskip \baselineskip
\caption{Moving soliton shuttle along the right half of the horizontal leg $(0<x<a)$ in the four-terminal geometry shown in Fig. 1c.  
$J(t)$ was instantaneously turned on from 0 to $1.012J_t$ at $t=0$, and the rest is the same as in Fig. 2.}
\label{fig.4}
\end{figure}


\begin{references}
\bibitem{imb}
If $\Delta_1$ appears at a higher $T_{c1}$, the induced
gap $\Delta_2=\gamma\Delta_1\cos\theta/\alpha_2$ at $T_{c2}<T<T_{c1}$ depends on $\theta$.
In that case $\theta$ is a poor variable, because Eq. (\ref{r}) turns into  
a single TDGL equation for $\Delta_1$ with $\alpha_1\to\alpha_1-\gamma^2\cos^2\theta/\alpha_2$, 
which describes only locked states with $\theta=0$ or $\pi$.
\bibitem{stat}
For $\tau_e\gg\tau_\theta$ and $L_e^2\gg L_\theta^2$, the ac component of
E obeys $\tau_e{\dot E}=L_e^2E''+\alpha_e{\dot\theta'}$, whence
$E'\simeq -\alpha_e{\dot\theta}/L_e^2$ for small soliton velocities $v\ll L_e/\tau_e$, and
$E'\simeq \alpha_e\theta''/\tau_e$ for $v\gg L_e/\tau_e$. This $E'$
just renormalizes the length and time scales in Eq. (\ref{theta}): $\tau_\theta\to\tau_\theta+\alpha_e\alpha_\theta/L_e^2$,
$L_\theta\to L_\theta$ for $v\ll L_e/\tau_e$,
and $\tau_\theta\to\tau_\theta$, $L_\theta^2\to L_\theta^2+\alpha_e\alpha_\theta/\tau_\theta$ for
$v\gg L_e/\tau_e$, but does not change the behavior of $\theta(x,t)$.
\end{references}
\end{document}